\documentclass[twocolumn,showpacs,amsmath,amssymb]{revtex4}

\usepackage{graphicx}
\usepackage{dcolumn}
\usepackage{bm}

\begin{document}

\title{Near-threshold $\Lambda$(1520) production by the $\vec{\gamma} p$ 
$\rightarrow$ $K^{+}\Lambda$(1520) reaction at forward $K^{+}$ angles}

\author{H.~Kohri$^{1}$, 
D.S.~Ahn$^{1,2}$, 
J.K.~Ahn$^{2}$, 
H.~Akimune$^{3}$, 
Y.~Asano$^{4}$,
W.C.~Chang$^{5}$, 
S.~Dat\'{e}$^{6}$, 
H.~Ejiri$^{1,6}$, 
S.~Fukui$^{7}$, 
H.~Fujimura$^{8,9}$, 
M.~Fujiwara$^{1,4}$, 
S.~Hasegawa$^{1}$, 
K.~Hicks$^{10}$, 
A.~Hosaka$^{1}$, 
T.~Hotta$^{1}$, 
K.~Imai$^{9}$, 
T.~Ishikawa$^{11}$, 
T.~Iwata$^{12}$, 
H.~Kawai$^{13}$, 
Z.Y.~Kim$^{8}$,
K.~Kino$^{1,a}$, 
N.~Kumagai$^{6}$, 
S.~Makino$^{14}$, 
T.~Matsuda$^{15}$, 
T.~Matsumura$^{16,1,4}$, 
N.~Matsuoka$^{1}$, 
T.~Mibe$^{1,4,10}$, 
M.~Miyabe$^{9}$,
Y.~Miyachi$^{17}$, 
M.~Morita$^{1}$, 
N.~Muramatsu$^{4,1}$, 
T.~Nakano$^{1}$, 
S.i.~Nam$^{18}$, 
M.~Niiyama$^{9}$, 
M.~Nomachi$^{19}$, 
Y.~Ohashi$^{6}$,  
H.~Ohkuma$^{6}$, 
T.~Ooba$^{13}$, 
D.S.~Oshuev$^{5,b}$, 
C.~Rangacharyulu$^{20}$, 
A.~Sakaguchi$^{19}$, 
T.~Sasaki$^{9}$, 
P.M.~Shagin$^{21}$, 
Y.~Shiino$^{13}$, 
A.~Shimizu$^{1}$, 
H.~Shimizu$^{11}$, 
Y.~Sugaya$^{19}$, 
M.~Sumihama$^{19,4}$, 
A.I. Titov$^{22}$
Y.~Toi$^{15}$, 
H.~Toyokawa$^{6}$, 
A.~Wakai$^{23}$, 
C.W.~Wang$^{5}$, 
S.C.~Wang$^{5}$, 
K.~Yonehara$^{3,c}$, 
T.~Yorita$^{1,6}$, 
M.~Yoshimura$^{24}$, 
M.~Yosoi$^{9,1}$, 
and R.G.T.~Zegers$^{25}$\\
(LEPS Collaboration)\\}

\affiliation{$^{1}$Research Center for Nuclear Physics, Osaka 
University, Ibaraki, Osaka 567-0047, Japan}
\affiliation{$^{2}$Department of Physics, Pusan National University, 
Busan 609-735, Korea}
\affiliation{$^{3}$Department of Physics, Konan University, 
Kobe, Hyogo 658-8501, Japan}
\affiliation{$^{4}$Kansai Photon Science Institute, Japan Atomic Energy 
Agency, Kizu, Kyoto 619-0215, Japan}
\affiliation{$^{5}$Institute of Physics, Academia Sinica, Taipei 11529, 
Taiwan}
\affiliation{$^{6}$Japan Synchrotron Radiation Research Institute, 
Mikazuki, Hyogo 679-5198, Japan}
\affiliation{$^{7}$Department of Physics and Astrophysics, Nagoya University, 
Nagoya, Aichi 464-8602, Japan}
\affiliation{$^{8}$School of Physics, Seoul National University, Seoul, 
151-747, Korea}
\affiliation{$^{9}$Department of Physics, Kyoto University, 
Kyoto 606-8502, Japan} 
\affiliation{$^{10}$Department of Physics And Astronomy, Ohio University, 
Athens, Ohio 45701, USA}
\affiliation{$^{11}$Laboratory of Nuclear Science, Tohoku University, 
Sendai, Miyagi 982-0826, Japan}
\affiliation{$^{12}$Department of Physics, Yamagata University, 
Yamagata 990-8560, Japan}
\affiliation{$^{13}$Department of Physics, Chiba University, 
Chiba 263-8522, Japan}
\affiliation{$^{14}$Wakayama Medical University, Wakayama, 
Wakayama 641-8509, Japan}
\affiliation{$^{15}$Department of Applied Physics, Miyazaki University, 
Miyazaki 889-2192, Japan}
\affiliation{$^{16}$Department of Applied Physics, National Defense 
Academy, Yokosuka 239-8686, Japan}
\affiliation{$^{17}$Department of Physics, Tokyo Institute of Technology, 
Tokyo 152-8551, Japan} 
\affiliation{$^{18}$Department of Physics, Chung-Yuan Christian 
University, Chung-Li 32023, Taiwan}
\affiliation{$^{19}$Department of Physics, Osaka University, Toyonaka, 
Osaka 560-0043, Japan}
\affiliation{$^{20}$Department of Physics, 
University of Saskatchewan, Saskatoon, Saskatchewan, Canada} 
\affiliation{$^{21}$School of Physics and Astronomy, University of Minnesota, 
Minneapolis, Minnesota 55455, USA}
\affiliation{$^{22}$Joint Institute of Nuclear Research, 141980, Dubna, 
Russia} 
\affiliation{$^{23}$Akita Research Institute of Brain and Blood Vessels, 
Akita 010-0874, Japan}
\affiliation{$^{24}$Institute for Protein Research, Osaka University, 
Suita, Osaka 565-0871, Japan}
\affiliation{$^{25}$National Superconducting Cyclotron Laboratory, 
Michigan State University, MI 48824, USA}

\date{\today}
\begin{abstract}
Differential cross sections and photon-beam asymmetries 
for the $\vec{\gamma} p$ $\rightarrow$ $K^{+}\Lambda$(1520) reaction
have been measured with linearly polarized 
photon beams at energies from the threshold to 2.4 GeV 
at 0.6$<\cos\theta_{\rm CM}^{K}<$1. 
A new bump structure was found at $W\simeq 2.11$ GeV in the 
cross sections. 
The bump is not well reproduced by theoretical calculations 
introducing a nucleon resonance with $J\leq\frac{3}{2}$. 
This result suggests that the bump might be produced by a 
nucleon resonance possibly with $J\geq\frac{5}{2}$ or by a new reaction 
process, for example an interference effect with the $\phi$ 
photoproduction having a similar bump structure 
in the cross sections. 
\end{abstract}

\maketitle


Strangeness photoproduction is an important tool to gain a deeper 
understanding of the nature of baryon resonances. 
Theoretically, constituent quark models predict more nucleon resonances 
than those observed in pion scattering reactions. 
Quark model studies suggest that these missing resonances couple to 
strangeness channels which are not only $KY$ ($Y$=$\Lambda$ or $\Sigma$) 
but also $KY^{*}$ ($Y^{*}$=$\Lambda^{*}$ or $\Sigma^{*}$)~\cite{Capstick}. 
Some nucleon resonances have been observed at the near-threshold 
energies in the $KY$ photoproduction \cite{Glander,Mcnabb,Sumihama}. 
The threshold for the $KY^{*}$ photoproduction is relatively high 
compared with that for the $\pi N$, $\eta N$, and $K Y$ photoproduction. 
Therefore, photoproduction leading to the $KY^{*}$ state is a 
good way to investigate poorly understood nucleon 
resonances with a heavy mass. 

Another physics interest in the $KY^{*}$ reaction is that the bump structure 
found at $E_{\gamma}\sim$2 GeV in the cross sections 
for the $\phi$ photoproduction~\cite{Mibe} 
might be explained by the coupled-channel or 
interference effects with relevant reactions~\cite{Ozaki1}. 
The cause of the bump has not been clarified yet. 
Measuring cross sections and spin observables for these relevant reactions, 
which have similar energy thresholds and final states, could  
play an important role in clarifying the cause of the bump. 
The $K^{+}\Lambda$(1520) photoproduction 
is one of the best reactions to satisfy 
the requirements for such a study. 

The reaction mechanism of the $K^{+}\Lambda$(1520) photoproduction is often 
described in terms of hadron exchanges, with $N$ and $N^{*}$ 
in the $s$-channel, $Y$ and $Y^{*}$ in the $u$-channel, 
and $K$ and $K^{*}$ in the $t$-channel. 
Recent theoretical studies suggest that the contact term (to satisfy 
the gauge invariance) is dominant and 
the $s$-channel contribution is negligibly small in the 
$K^{+}\Lambda$(1520) reaction \cite{Nam1,Nam2}. 
Another theoretical study suggests that the $K^{*}$ exchange 
contribution is small \cite{Toki}. 
On the other hand, previous $K^{+}\Lambda$(1520) photoproduction data 
at the center-of-mass (CM) energies ($W$=$\sqrt{s}$) of $W$=2.48-3.14 GeV 
($E_{\gamma}$=2.8-4.8 GeV) show that $K^{*}$ exchange 
in the $t$-channel is dominant \cite{Barber}. 
Recent $K^{+}\Lambda$(1520) electroproduction data at 
$W$=1.95-2.65 GeV show that contributions from $K$ and $K^{*}$ 
exchanges are roughly equal \cite{Barrow}. 
Therefore, additional data with new observables are needed 
for solving this controversial situation. 
The photon-beam asymmetry ($\Sigma$) for $K^{+}\Lambda$(1520) 
photoproduction has some unique features. 
Nam $et$ $al$. predict that 
$\Sigma=-1$ or $\Sigma >$0 if the $K$ or $K^{*}$ meson 
is exchanged in the $t$-channel, respectively \cite{Nam2}. 
The contact term, $u$-channel, and $s$-channel $N$ exchange 
contributions give almost zero asymmetries. 
Hence, a measurement of the $\Sigma$ asymmetry provides strong 
constraints in understanding the $K^{+}\Lambda$(1520) 
photoproduction mechanism. 

In the past, experimental data for hyperon photoproduction 
at the near-threshold energies were available only for 
the $K^{+}\Lambda$ and $K^{+}\Sigma^{0}$ states
\cite{Glander,Mcnabb,Zegers,Sumihama,Bradford}. 
Recently, new experimental results for $K^{0}\Sigma^{+}$ \cite{Castelijns}, 
$K^{+}\Sigma^{-}$ \cite{Kohri}, $K^{*0}\Sigma^{+}$ \cite{Hleiqawi,Nanova}, 
$K^{+}\Lambda$(1405) \cite{Niiyama}, 
and $K^{+}\Sigma^{-}$(1385) \cite{Hicks} have been reported. 
However, there are only two old published results on $K^{+}\Lambda$(1520) 
photoproduction at energies of $E_{\gamma}$=2.8-4.8 GeV \cite{Barber} 
and $E_{\gamma}$=11 GeV \cite{Boyarski2}. 
New experimental data near the $K^{+}\Lambda$(1520) 
threshold are useful 
to investigate the possibility of new nucleon resonances, 
to obtain key information for clarifying the cause of the bump found 
in the $\phi$ photoproduction, 
and to understand the $K^{+}\Lambda$(1520) reaction mechanism. 
In this Letter, we present, for the first time, 
differential cross sections and photon-beam asymmetries for the 
$\vec{\gamma} p$ $\rightarrow$ $K^{+}\Lambda$(1520) reaction 
at 0.6$<\cos\theta_{\rm CM}^{K}<$1 at the near-threshold energies. 


The experiment was carried out using the laser-electron 
photon facility at SPring-8 (LEPS) \cite{Nakano}. 
The energy range of tagged photons was 1.5-2.4 GeV, and 
the polarization of linearly polarized photons 
was 52-90\% at 1.5-2.4 GeV. 
We used a liquid hydrogen (LH$_{2}$) target with 
an effective length of 16 cm. 
Charged particles produced at the target were detected at forward angles 
with the LEPS spectrometer system for trajectory tracking. 
Time-of-flight information was obtained for each charged particle 
track. 
The start signal was produced by a plastic scintillator (SC) located 
behind the target, and the stop signal was produced by an array of 40 
plastic scintillators at the downstream of the spectrometer. 
The $K^{+}$ meson was identified from its mass, within 3$\sigma$ 
where $\sigma$ is the momentum dependent mass resolution. 
The data sample with the single $K^{+}$ meson was analyzed. 

\begin{figure}[h]
\begin{center}
\includegraphics[width=7.5 cm,height=4.0 cm]{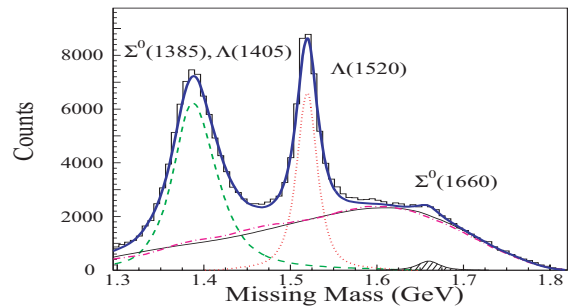}%
\caption{\label{fig_1} Missing mass of the $p$($\gamma$,$K^{+}$)$X$ 
reaction at $E_{\gamma}$=1.5-2.4 GeV 
and 0.6$<\cos\theta_{\rm CM}^{K}<$1.
The thick solid curve is the result of the fit using the polynomial 
background (thin solid curve). 
The dashed, dotted, and hatched curves correspond to 
$\Sigma^{0}$(1385)/$\Lambda$(1405), $\Lambda$(1520), and 
$\Sigma^{0}$(1660) productions, respectively. 
The dotted-dashed curve is the background obtained by the fit 
using simulation curves. }
\label{fig:missall} 
\end{center}
\end{figure}

Figure \ref{fig:missall} shows the missing mass ($MM_{\gamma K^{+}}$) 
spectrum for the $p$($\vec{\gamma}$,$K^{+}$)$X$ 
reaction. 
The wide lower-mass peak corresponds to the $\Sigma^{0}$(1385) and 
$\Lambda$(1405) production, and the narrow higher-mass peak 
corresponds to the $\Lambda$(1520). 
The $\Lambda$(1520) yield was obtained by fitting the peaks 
in the missing mass spectrum. 
The photon energy region from the threshold to 2.4 GeV was divided 
into 15 bins and the $K^{+}$ polar angle region in the CM system 
was divided into 4 bins. 
The peak shape of each hyperon resonance was estimated 
by GEANT simulations. 
Breit-Wigner shapes with masses of 1.384, 1.407, and 1.520 GeV 
and widths of 36, 50, and 16 MeV 
were used to generate the $\Sigma^{0}$(1385), $\Lambda$(1405), and 
$\Lambda$(1520) hyperon resonances, respectively \cite{PDG}. 
The masses and widths of the hyperon resonances are 
uncertain \cite{PDG}, and
these uncertainties were evaluated as systematic errors. 
The peak shape was reproduced by the missing mass of 
the $p$($\gamma$,$K^{+}$)$X$ reaction in the simulations 
including the experimental resolution. 
The peak shape was fixed in the fit to the experimental missing mass 
spectrum and the height of the peak was adjusted as a free parameter. 
There is a small bump at 1.66 GeV, probably due to the $\Sigma^{0}$(1660).
Since the mass and width of the $\Sigma^{0}$(1660) are not well known, 
the same peak shape as the $\Lambda$(1520) was used, but with its position
fixed at 1.660 GeV in the fit. 

The background under the hyperon peaks was fit by using 
a polynomial function. 
The $\gamma p$ $\rightarrow$ $K^{+}\pi Y$, $K^{*}Y$, 
$K^{+}K N$, and $\phi p$ reactions account for the majority 
of the background under the hyperon peaks in the simulation studies. 
The $K^{+}\pi Y$ and $K^{*}Y$ reactions are considered to be dominant 
at $MM_{\gamma K^{+}}<$1.5 GeV, while the $\phi p$ reaction is 
dominant at $MM_{\gamma K^{+}}>$1.5 GeV. 
As a result of the fit, the $\Lambda$(1520) yield 
was obtained for each incident photon energy and angular bin. 
The differential cross sections for the $K^{+}\Lambda$(1520) reaction 
were obtained by using the same method in Ref.~\cite{Sumihama}. 


A fit with background curves generated for the 
$\gamma p$ $\rightarrow$ $K^{+}\pi Y$, $K^{*}Y$, 
$K^{+}K N$, and $\phi p$ reactions by the simulations 
makes a difference of at most 0.1 $\mu$b for the $K^{+}\Lambda$(1520) 
cross sections. 
The sum of the background curves is shown in Fig.~\ref{fig:missall}. 
Systematic uncertainties of the shape, mass, and width of the 
$\Lambda$(1520), $\Lambda$(1405), and $\Sigma^{0}$(1385) 
resonances cause uncertainties of 0.04 $\mu$b at $W<$2.15 GeV 
and 0.07 $\mu$b at $W>$2.15 GeV. 
Uncertainties of the target thickness, photon flux, and 
detector acceptance are 1\%, 5\%, and 3\%, respectively. 
The $\pi^{+}$ contamination in the particle identification of 
the $K^{+}$ is negligibly small. 
When the $K^{+}$ is detected at forward angles, the vertex 
resolution becomes poor. 
The contamination of events from the SC in the vertex selection of 
the LH$_{2}$ target is smaller than 3\% at $W>2.04$ GeV
and smaller than 7\% at $W<$2.04 GeV. 

\begin{figure}[htb]
\begin{center}
\includegraphics[width=7.7 cm,height=12.4 cm]{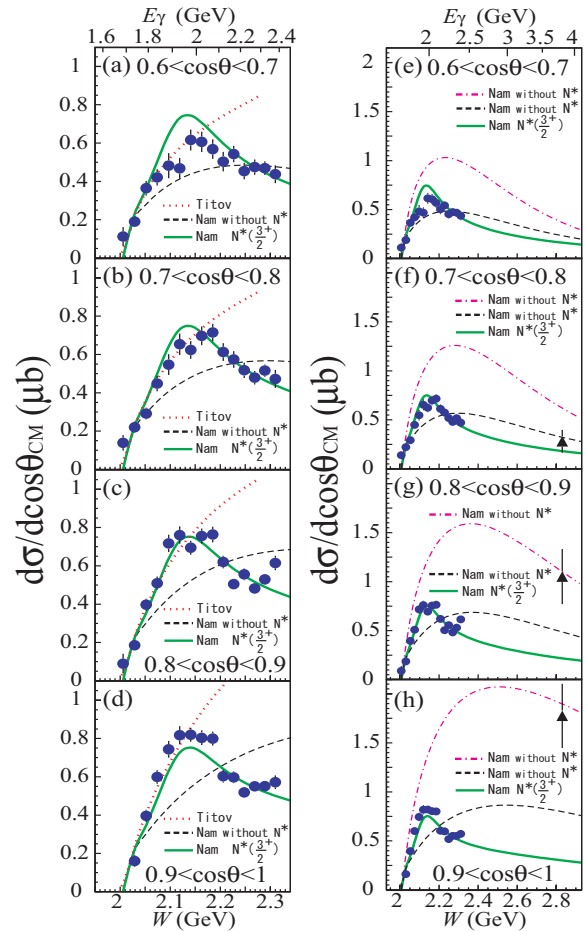}%
\caption{Differential cross sections 
for the 
$K^{+}\Lambda$(1520) reaction at 
(a, e) 0.6$<\cos\theta_{\rm CM}^{K}<$0.7, 
(b, f) 0.7$<\cos\theta_{\rm CM}^{K}<$0.8, 
(c, g) 0.8$<\cos\theta_{\rm CM}^{K}<$0.9, and 
(d, h) 0.9$<\cos\theta_{\rm CM}^{K}<$1. 
The circles are the present data. 
The circles in the left and right figures are the same data. 
The triangles are the Daresbury data ($E_{\gamma}$=2.8-4.8 GeV)~\cite{Barber}. 
The solid and dashed curves are the results of calculations fitting 
to the present data by Nam $et$ $al$. with and without 
a nucleon resonance($J^{\pi}$=$\frac{3}{2}^{+}$), 
respectively~\cite{Nam3}. 
The dotted-dashed curves are the results of calculations fitting to 
the Daresbury data by Nam $et$ $al$. \cite{Nam1}. 
The dotted curves are the results of calculations by 
Titov $et$ $al$. \cite{Titov}. }
\label{fig:cross}
\end{center}
\end{figure}

The differential cross sections for the 
$\vec{\gamma} p$ $\rightarrow$ $K^{+}\Lambda$(1520) reaction 
are shown in Fig.~\ref{fig:cross}. 
The cross sections increase with the CM energy 
near the threshold. 
It is quite interesting that the experimental cross sections rapidly 
decrease at around $W$=2.2 GeV and a clear bump structure is 
observed at the $K^{+}$ angles of 0.8$<\cos\theta_{\rm CM}^{K}$. 
The rapid decrease at around $W$=2.2 GeV is much larger 
than the statistical and systematic errors. 
This bump energy is similar to the energy where another bump was found 
in the $\phi$ photoproduction~\cite{Mibe}. 
Note that the bump at this energy is not observed in the 
$K^{+}\Lambda$(1116) \cite{Sumihama}, 
$K^{+}\Sigma^{0}$ \cite{Sumihama,Kohri}, 
$K^{+}\Sigma^{-}$ \cite{Kohri}, or $K^{+}\Sigma^{-}$(1385) \cite{Hicks} 
cross sections obtained using the same method. 

\begin{figure}[htb]
\begin{center}
\includegraphics[width=7.8 cm,height=3.7 cm]{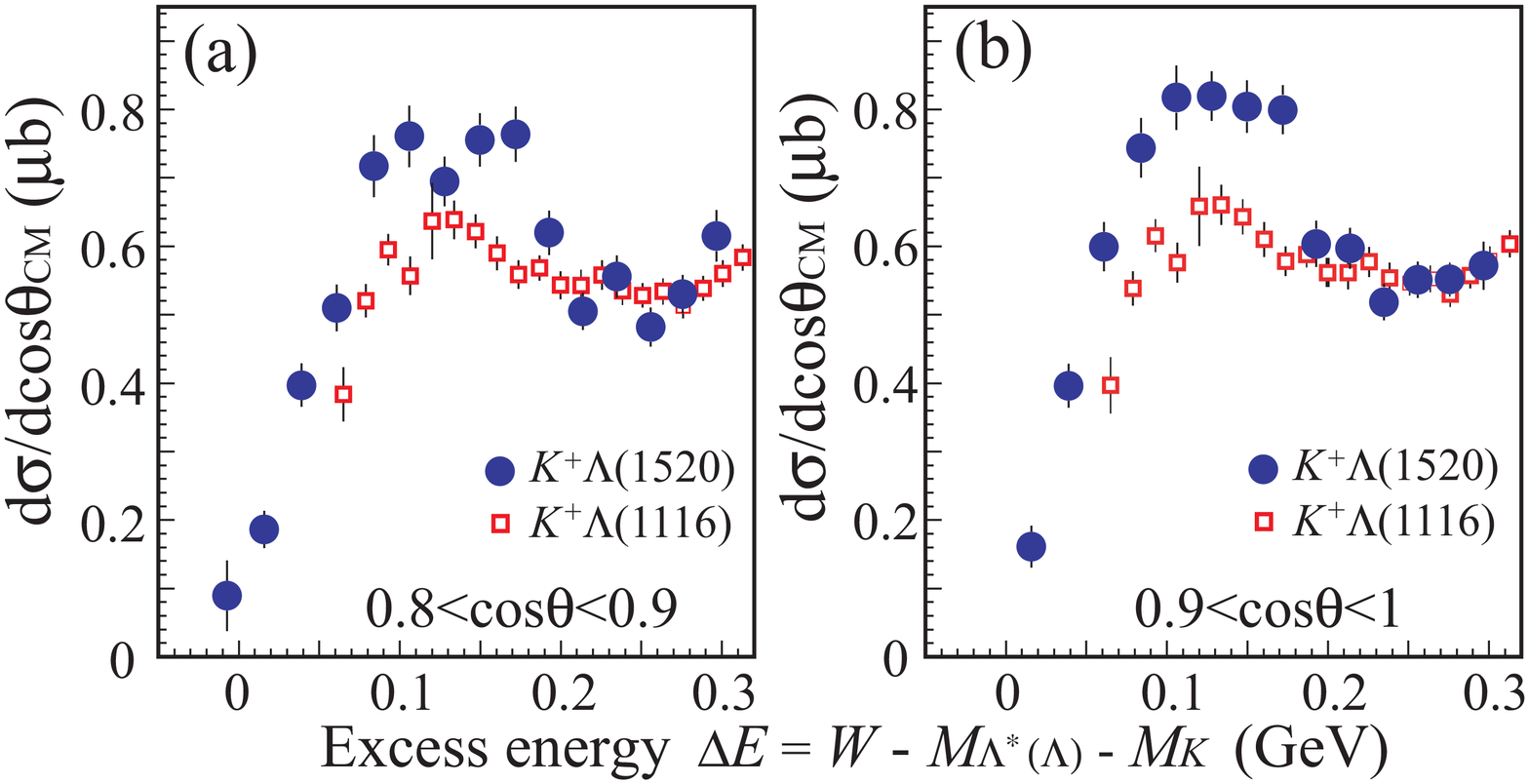}%
\caption{Differential cross sections 
for the $K^{+}\Lambda$(1520)(circles) reaction at 
(a) 0.8$<\cos\theta_{\rm CM}^{K}<$0.9 and 
(b) 0.9$<\cos\theta_{\rm CM}^{K}<$1 as a function of 
the excess energy ($\Delta E$). 
The squares are prescaled differential cross sections 
for the $K^{+}\Lambda$(1116) reaction
at 0.8$<\cos\theta_{\rm CM}^{K}<$0.9~\cite{Mcnabb}. 
The prescale factors of 0.263 for (a) and 0.271 for (b) are 
used to fit the $K^{+}\Lambda$(1116) cross sections to 
the $K^{+}\Lambda$(1520) cross sections at 0.2 GeV$<$ $\Delta E$ $<$0.3 
GeV. 
There are no $K^{+}\Lambda$(1116) data at 0.9$<\cos\theta_{\rm CM}^{K}<$1.}
\label{fig:eecross}
\end{center}
\end{figure}

The $K^{+}\Lambda$(1520) cross sections are compared with the prescaled 
$K^{+}\Lambda$(1116) cross sections~\cite{Mcnabb} as a function of the 
excess energy in Fig.~\ref{fig:eecross}. 
The clear bump structure found in the present $K^{+}\Lambda$(1520) cross 
sections is not seen in the forward-angle $K^{+}\Lambda$(1116) cross sections, 
which suggests that the reaction mechanism is different between 
the two reactions at these near-threshold energies. 

Two theoretical calculations, which are based on an effective 
Lagrangian approach, by Titov $et$ $al$. \cite{Titov} and 
Nam $et$ $al$. \cite{Nam1} monotonically increase 
with the CM energy up to $W\sim$2.3 GeV in 
Fig.~\ref{fig:cross}(a, b, c, d). 
The calculations by Titov $et$ $al$. are not tuned to 
fit the data. 
The calculations by Nam $et$ $al$. are dominated by the contact 
term contribution. 
Although the results of the calculations by Nam $et$ $al$. approach 
the present data by optimizing the cutoff parameter, 
the rapid decrease associated with the bump cannot be reproduced. 
The agreement with the present data is poor. 

As one possibility, we perform new calculations to describe the present 
data by introducing a nucleon resonance with a free mass and 
a width~\cite{Nam3}, although the angular coverage of the data is 
inadequate to obtain strong evidence for the nucleon resonance. 
The spins and parities of $J^{\pi}=\frac{1}{2}^{\pm}$ and 
$\frac{3}{2}^{\pm}$ for the nucleon resonance are considered. 
Contributions from the nucleon resonance with a spin higher than 
$\frac{3}{2}$ are not included due to theoretical ambiguities. 
The angular distributions of the $J^{\pi}=\frac{1}{2}^{\pm}$ and 
$\frac{3}{2}^{\pm}$ states are almost flat. 
The $J^{\pi}=\frac{3}{2}^{+}$ state gives a better 
reduced $\chi^{2}$(1.37) for the fit than the other states, and 
the energy dependence of the bump is reproduced 
by the solid curves of Fig.~\ref{fig:cross}. 
However, the angular distribution of the bump is not well reproduced. 
The theoretical calculations estimate the cross sections 
at backward $K^{+}$ angles to be about 0.7 $\mu$b that 
overestimates the experimental cross sections~\cite{Muramatsu} 
by 2-3 times. 
The bump is not observed in the cross sections 
of the backward $K^{+}$ angles~\cite{Muramatsu}. 

As a result of the fit, the mass and width of the 
$J^{\pi}=\frac{3}{2}^{+}$ nucleon resonance are obtained 
as 2.11 GeV and 140 MeV, respectively. 
Nucleon resonances with similar masses are $D_{13}$(2080) 
with 2-star status, $S_{11}$(2090) and 
$P_{11}$(2100) with 1-star status, and 
$G_{17}$(2190) with 4-star status in the PDG particle listings \cite{PDG}. 
In the listings, there is no corresponding nucleon resonance at 2.11 GeV. 
Note that most of the widths measured for the nucleon resonances in 
the listings are much wider than 140 MeV. 
Quark model studies predict that a new $J^{\pi}=\frac{5}{2}^{-}$ state 
with a similar mass of 2.08 GeV may be visible in the 
$K^{+}\Lambda$(1520) reaction \cite{Capstick}. 
Theoretical improvements for introducing a nucleon resonance 
with spins higher than $\frac{3}{2}$ are important to judge whether 
the bump is produced by the nucleon resonance or not. 

Another possible explanation for the bump is a new reaction process, 
for example, an interference effect between the $\phi$ and 
$\Lambda$(1520) photoproduction reactions might produce the bump 
because both reactions have this feature in the cross sections 
at similar energies \cite{Mibe}. 
Coupled-channel effects are unlikely to reproduce the strength 
and the angular distribution of the bump~\cite{Ozaki2}. 

Typical cross sections for hyperon photoproduction, such as 
$K^{+}\Lambda$(1116) and $K^{+}\Sigma^0$(1193), show a gradual decrease
with increasing the CM energy \cite{Bradford}. 
A gradual decrease in the $K^{+}\Lambda$(1520) cross sections 
\cite{Barber} is reproduced by the calculations of 
Nam $et$ $al$.~\cite{Nam1,Nam3} as shown by all the 
curves of Fig.~\ref{fig:cross}(e, f, g, h). 
Although the connection between the present data and the Daresbury data 
seems to be smooth at 0.7$<\cos\theta_{\rm CM}^{K}<$0.8, 
the Daresbury data at 0.8$<\cos\theta_{\rm CM}^{K}$ are not smoothly 
extrapolated from the present data as shown by the solid curves of 
Fig.~\ref{fig:cross}(f, g, h). 
The differences between the Daresbury data and the solid curves 
are larger than three standard deviations at 0.8$<\cos\theta_{\rm CM}^{K}$. 
Calculations that would agree well with both data sets are very 
difficult at present. 
New experimental data ($W$=2.3-2.8 GeV) 
are desired to fill the gap between these two data sets. 


By using vertically and horizontally polarized photon 
beams, the photon-beam asymmetry has been shown to be insensitive
to the spectrometer acceptance \cite{Zegers,Sumihama}. 
The asymmetry ($\Sigma$) is given as  
$P_{\gamma}\Sigma$ $\cos2\phi$ = ($N_{v}-N_{h}$)/($N_{v}+N_{h}$), 
where $N_{v}$ and $N_{h}$ are the $\Lambda$(1520) yields 
with the vertically and horizontally polarized photons, respectively. 
$P_{\gamma}$ is the polarization degree of the photon beam, and 
$\phi$ is the $K^{+}$ azimuthal angle defined by the angle between 
the reaction plane and the horizontal plane. 
The photon energy region from the threshold to 2.4 GeV was 
divided into 7 bins and the $K^{+}$ azimuthal angle region 
was divided into 9 bins. 
The $K^{+}$ polar angle region was not divided. 
The $\Lambda$(1520) yields were obtained for each energy 
and angular bin by fits to the missing mass. 

\begin{figure}[htb]
\includegraphics[width=7.8 cm,height=3.6 cm]{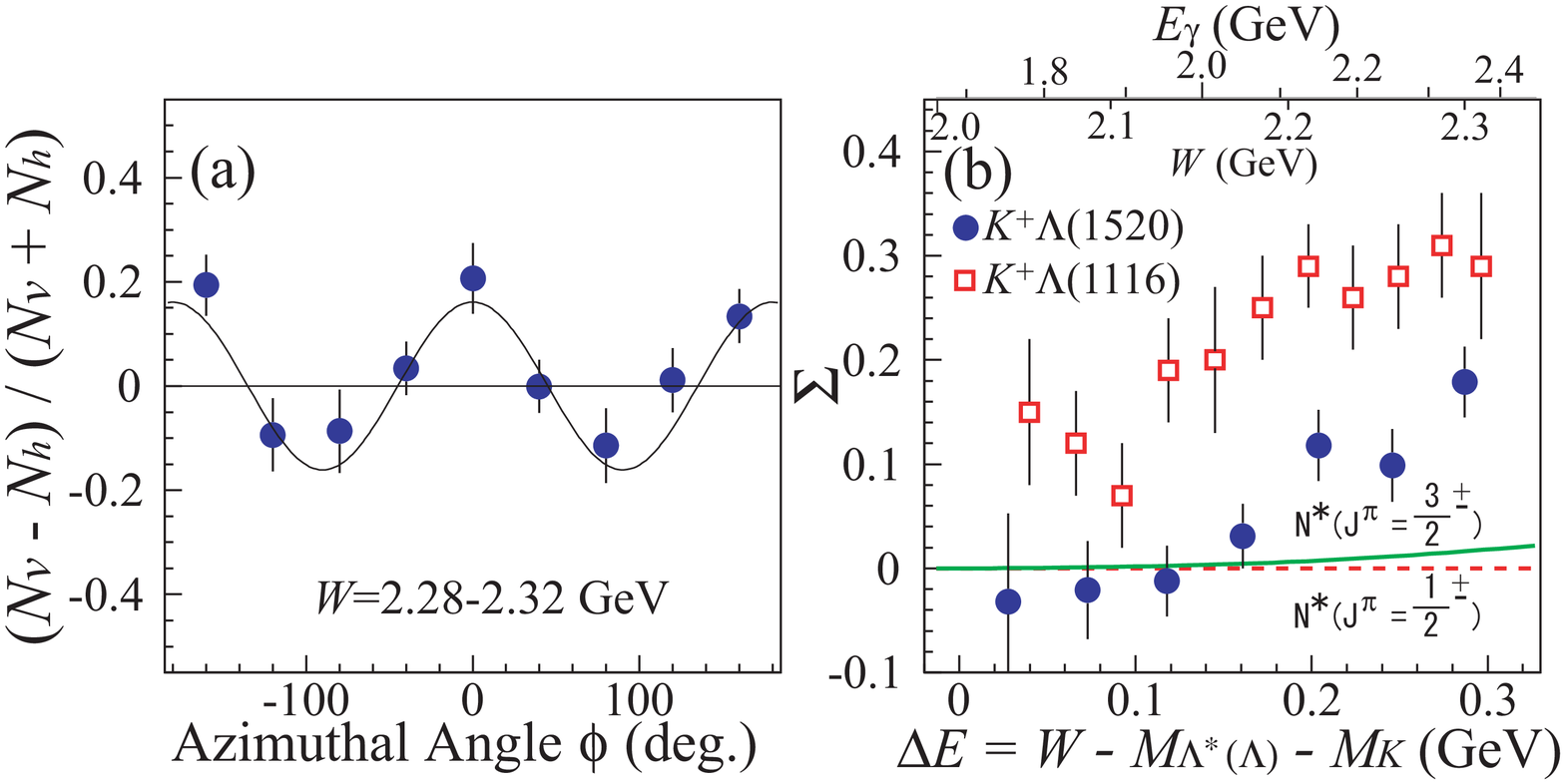}%
\caption{\small (a) Azimuthal angle distribution of the 
ratio ($N_{v}-N_{h}$)/($N_{v}+N_{h}$) for the 
$K^{+}\Lambda$(1520) reaction at $W$=2.28-2.32 GeV. 
(b) Photon-beam asymmetries for the 
$K^{+}\Lambda$(1520)(circles) and $K^{+}\Lambda$(1116)(squares)~\cite{Lleres} 
reactions as a function of the excess energy ($\Delta E$). 
The dashed and solid curves 
are the results of calculations for the $K^{+}\Lambda$(1520) 
by Nam $et$ $al$. introducing the $J^{\pi}=\frac{1}{2}^{\pm}$ 
and $J^{\pi}=\frac{3}{2}^{\pm}$ 
nucleon resonances, respectively~\cite{Nam3}. 
The parity for the resonances does not change the theoretical 
asymmetries significantly. 
The result of calculations without a nucleon resonance is 
almost identical to the solid curve.} 
\label{fig:ba}
\end{figure}

Figure \ref{fig:ba}(a) shows the $K^{+}$ azimuthal angle 
distribution of the ratio ($N_{v}-N_{h}$)/($N_{v}+N_{h}$) 
at $W$=2.28-2.32 GeV. 
The amplitude of the fit curve was divided by $P_{\gamma}$ and 
the asymmetry $\Sigma$ was obtained. 
Systematic uncertainties of the shape, mass, and width of the 
$\Lambda$(1520), $\Sigma^{0}$(1385), $\Lambda$(1405), and 
$\Sigma^{0}$(1660) hyperon resonances cause the uncertainty, 
$\delta\Sigma$=0.05. 
A fit with background curves generated for the 
$\gamma p$ $\rightarrow$ $K^{+}\pi Y$, $K^{*}Y$, 
$K^{+}K N$, and $\phi p$ reactions by the simulations 
is consistent within the statistical error. 
The effect of the $\pi^{+}$ contamination in the $K^{+}$ selection 
is negligible. 
When the $K^{+}$ is detected at forward angles, the vertex 
resolution becomes poor. 
The effect of the contamination of events from the SC 
in the vertex selection of the LH$_{2}$ target is also negligible. 
The attenuation of the asymmetry by the finite number of the 
azimuthal angle bins (9 bins) is about $\delta\Sigma$=0.015. 
The systematic uncertainty of the measurement of the laser 
polarization is $\delta\Sigma$=0.02. 

Figure~\ref{fig:ba}(b) shows the photon-beam asymmetries 
for the $K^{+}\Lambda$(1520) reaction at 
0.6$<\cos\theta_{\rm CM}^{K}<$1 in comparison with those 
for the $K^{+}\Lambda$(1116) reaction 
at $\cos\theta_{\rm CM}^{K}\sim$0.85~\cite{Lleres}. 
The $K^{+}\Lambda$(1520) asymmetries are near zero 
at $W<$2.2 GeV and increase gradually with the CM energy. 
The small positive values at $W>$2.2 GeV might indicate that 
the contribution from the $K^{*}$ exchange is larger than that 
from the $K$ exchange. 
The asymmetries for the $K^{+}\Lambda$(1520) reaction are smaller 
than those for the $K^{+}\Lambda$(1116) reaction. 
One reason of the small $K^{+}\Lambda$(1520) asymmetries is that 
the $K^{*}$ exchange contribution 
may be smaller than that in the $K^{+}\Lambda$(1116) reaction. 
The contact term and $K$ exchange contributions make the 
$K^{+}\Lambda$(1520) asymmetries smaller. 
This comparison suggests that the $K^{+}\Lambda$(1520) reaction 
mechanism is different from the $K^{+}\Lambda$(1116) reaction mechanism 
at these near-threshold energies. 

The $K^{+}\Lambda$(1520) asymmetry data are compared with the results 
of theoretical calculations by Nam $et$ $al$. with and without 
a nucleon resonance ($J^{\pi}=\frac{1}{2}^{\pm}$ or 
$\frac{3}{2}^{\pm}$) for the bump~\cite{Nam3}. 
The calculations use the parameters obtained from fits to the 
present cross sections. 
There is no significant difference between the results of these 
calculations as shown in Fig.~\ref{fig:ba}(b). 
Since all theoretical asymmetries are close to zero and agree with 
the data at $W<$2.2 GeV, we cannot judge whether the bump is 
due to a nucleon resonance or not. 
The measurement of additional spin observables is needed to 
clarify the cause of the bump. 
The calculations underestimate the data by 1-3 standard deviations 
including the systematic uncertainties above the bump energy. 
The contribution from the $K^{*}$ exchange is estimated to be larger 
than that obtained from fits to just the cross section data. 


In summary, we have measured differential cross sections and 
photon-beam asymmetries for 
the $\vec{\gamma} p$ $\rightarrow$ $K^{+}\Lambda$(1520) 
reaction. 
A bump structure was found in the cross sections. 
As one possible explanation, we introduce a nucleon 
resonance with $J\leq\frac{3}{2}$ 
in the theoretical calculations dominated by 
the contact term contribution, although the angular coverage of our 
data is inadequate to obtain strong evidence for the nucleon resonance. 
The calculations reproduce the energy dependence of the bump 
at forward $K^{+}$ angles, but 
fail to reproduce the angular distribution of the bump. 
Further theoretical calculations with $J\geq\frac{5}{2}$ resonances 
are necessary to examine the presence of a nucleon resonance. 
Another possible explanation is that 
the bump might be produced by a new reaction process, for example 
an interference effect with $\phi$ photoproduction. 
The $K^{+}\Lambda$(1520) asymmetries have small positive values 
at $W>$2.2 GeV, which may indicate that the contribution from 
the $K^{*}$ exchange is 
larger than that from the $K$ exchange. 
The asymmetries for the $K^{+}\Lambda$(1520) reaction are 
smaller than those for the $K^{+}\Lambda$(1116) reaction, which confirms that 
the $K^{+}\Lambda$(1520) reaction mechanism is different from 
the $K^{+}\Lambda$(1116) reaction mechanism 
at these near-threshold energies. 
The present result stimulates future experimental and theoretical 
studies for not only the $KY^{*}$ photoproduction reaction 
but also other relevant reactions with wider angular coverage, and 
will advance our understanding 
of the hadron photoproduction and baryon resonance. 

The authors thank the SPring-8 staff for supporting the 
experiment. 
We thank Mr. S. Ozaki for fruitful discussions. 
This research was supported in part by the Ministry of 
Education, Science, Sports and Culture of Japan, by 
the National Science Council of Republic of China (Taiwan), 
by the National Research Foundation of Korea, 
and by National Science Foundation (USA). 

\begin{small}
\hspace*{-0.4cm}$^{a}$Present address: Faculty and Graduate School of 
Engineering, Hokkaido University, Sapporo 060-8628, Japan. \\
$^{b}$Present address: Nuclear Physics Institute, Moscow State 
University, Moscow, 119899, Russia. \\
$^{c}$Present address: Illinois Institute of Technology, 
Chicago, IL 60616, USA.\\
\end{small}

\end{document}